\documentclass[12pt,a4paper]{article}

\usepackage{epsfig,amsmath,amssymb}

\def\wdw{Wheeler--DeWitt}
\def\sch{Schr\"{o}dinger}
\def\be{\begin{equation}}
\def\ee{\end{equation}}

\def\tp1{\tilde p_1}
\def\tp2{\tilde p_2}
\def\tq1{\tilde q^1}
\def\tq2{\tilde q^2}
\def\om{\Omega}

\begin{document}

\baselineskip0.27in

\title{Comment on the choice of time in a two-component formulation of the Wheeler--DeWitt equation} 
\author{Marc Thibeault\thanks{e-mail: marc@df.uba.ar}\ \ and
Claudio Simeone\thanks{e-mail: csimeone@df.uba.ar}\\
 {\small Departamento de F\'{\i}sica, Facultad de Ciencias Exactas y 
Naturales,}\\ 
{\small Universidad de Buenos Aires, 
Ciudad Universitaria
 Pab. I,}\\ 
{\small 1428, 
Buenos Aires, Argentina}}

\maketitle

\begin{abstract}
The two-component formalism in quantum cosmology is revisited with a particular
 emphasis on the identification of time. Its relation with the appearance of 
imaginary eigenvalues is established. It is explicitly shown how a good choice of the 
global time prevents  this peculiarity.

\end{abstract}

\bigskip


\bigskip


\vskip1cm

In Ref. \cite{mos98} an interesting proposal was introduced for solving the problem of the interpretation of the  \wdw\  equation \cite{ha,ba93} for minisuperspaces. The proposal is inspired in Dirac's solution to the problems of the
Klein--Gordon equation, and consists  in writing the  \wdw\  equation as a two-component \sch\
equation.
This  formulation is a possible way for associating a system of differential equations which are  first order in the derivative respect to time to a  Klein--Gordon type equation, as it is the Wheeler--DeWitt one. This allows to obtain  an   equation which admits  well known resolution procedures   and  an interpretation in terms of a well defined inner product.  In the program developed in Refs. \cite{mos98,mos0205,mos0209} it was shown that such idea can be  applied to  minisuperspace models. The procedure reduces the resolution of the Wheeler--DeWitt equation to an eigenvalue problem  and a series of algebraic equations which can be solved by iteration.
The proposal has been developed within the framework of the quantization of  Friedmann--Robertson--Walker (FRW) minisuperspaces. In the case of a closed ($k=1$) model with matter in the form of a massive scalar field, eigenvalues which become imaginary as the system evolves were obtained in one of the steps of the application of the formalism. Though the program of Refs. \cite{mos98,mos0205,mos0209} does not rely on  a standard Schr\"odinger equation, where imaginary eigenvalues lead to a non unitary quantum description, the point deserves, we believe, a closer look which goes beyond the particular models considered in \cite{mos98}. The possible relation existing between this  peculiarity  and the particular choice of time variable has been already pointed in Ref.  \cite{ss04}.  The purpose of the present note is to carry out a detailed analysis  of this issue  within the context of the search for a consistent minisuperspace quantization with a globally right notion of time.\\

For a FRW minisuperspace with scaled constraint of the form
\be
H=\eta^{\mu\nu}p_\mu p_\nu+V(q)=0\label{hh}
\ee
where $\eta^{\mu\nu}=\mathrm{diag}(-1,1)$ and $p$ are the momenta conjugated to the coordinates $q$, the program of Refs. \cite{mos98,mos0205,mos0209} starts with the identification of $q^0$ as time variable and the definition of the operator
\be
\hat D\equiv -\frac{\partial^2}{\partial (q^1)^2}+{\hat V}(q).\label{dd}
\ee
 Then the second order equation resulting from the Hamiltonian constraint   is reduced   defining a two-component wave function
\be
\Psi={1\over {\sqrt 2}}\left( \begin{array}{c}\psi+i\dot\psi \\ \psi-i\dot
\psi \end{array}
\right)
\ee
where dots mean derivatives respect to $q^0$, and introducing the time-dependent Hamiltonian operator
\be
\hat{H}={1\over 2}\left( \begin{array}{cc} 1+\hat D & -1+\hat D \\ 1-\hat D & -1-\hat D\end{array}
\right).
\ee
 This leads to the Schr\"{o}dinger   equation
\be
i\dot\Psi=\hat{H}\Psi
\ee
which is then solved by finding the solutions for the eigenvalue problem 
\be
\hat{H}\Psi_n=E_n\Psi_n.\label{hal}
\ee
 With the definition
\be
\Psi_n=\frac{1}{\sqrt{2}}\left( \begin{array}{c} 1+E_n \\ 1-E_n \end{array}
\right)\Phi_n
\ee
one obtains
\be
\hat D\Phi_n=E_n^2\Phi_n.\label{D}
\ee
 In Ref. \cite{mos98} the particular case of a model of arbitrary curvature with a massive scalar field $\phi$ was considered, with the choice $q^0=\Omega$,  so that  $\hat D= -{\partial^2\over\partial\phi^2}-ke^{4\Omega}+m^2\phi^2e^{6\Omega}$. Thus Eq. (\ref{D}) results to be
\be
\left\{-{\partial^2\over\partial\phi^2}+m^2\phi^2e^{6\Omega}\right\}\Phi_n={\cal E}_n\Phi_n
\ee 
where ${\cal E}_n=E_n^2+ke^{4\Omega}$ (see Ref. \cite{mos0205} for the details). In this case ${\cal E}_n$ are chosen as the energies of the harmonic oscillator. However,  for the closed $(k=1)$ model, the eigenvalues  $E_n=\sqrt{{\cal E}_n-e^{4\Omega}}$ become imaginary as $\Omega$ increases: this happens for  $e^\Omega>m$ (the corresponding eigenvectors are null).\\ 

Now, it is  known that for the closed FRW model  the logarithm of the scale factor  is not a globally good time variable since its  evolution is not monotonous (the model undergoes a re-collapse). A globally good time is a function $t(q^i,p_i)$ which monotonically
increases along a dynamical trajectory, that is, each surface
$t=\rm constant$ in the phase space is crossed by a dynamical
trajectory only once; hence the successive states of the system can
be parametrized by this function. This means that $t(q^i,p_i)$
must fulfill the condition
\cite{ha86}
\be
[t, H]>  0.
\ee
Because  the general form of the Hamiltonian constraint is $H=G^{ik}p_ip_k+V(q)=0$ and the supermetric $G^{ik}$ \cite{ha,ba93} does not depend on the  momenta, a
function of only the coordinates $t(q^i)$ is a global time if the bracket
$
 [t(q^i), G^{ik}p_ip_k]
 = 2 G^{ik}p_k\,\partial t/\partial q^i
 $
is positive definite. Hence if the supermetric has a diagonal form
and one of the momenta vanishes at a given point of phase space,
then no function of only its conjugated coordinate can be a global
time. For a constraint whose potential can be zero for finite
values of the coordinates, the momenta $p_k$ can be all equal to
zero at a given point, and $[t(q^i),H]$ can vanish. Hence, what is called an
{\it intrinsic time} $t(q^i)$ \cite{ku92} exists only if the potential in the
constraint has a definite sign; this is closely related with the existence of a globally well defined canonical gauge condition $\chi(q,\tau)=0$ for the system \cite{si99}, as it is to be expected from the fact that in the theory of gravitation the dynamical evolution can be reproduced by gauge transformations \cite{ba93}. 
In the most general case a global
phase time should be a function including the canonical momenta;
this is called an {\it extrinsic time} $t(q^i,p_i)$ \cite{ku71}.\\

To show explicitly the  relation of imaginary eigenvalues problem with a choice of time which is not globally right we shall consider  a model which, despite its simplicity, allows for a detailed analysis of the central aspects of the problem.
A suitable example  is
given by an isotropic and homogeneous dilaton cosmology described by the scaled
Hamiltonian constraint
\begin{equation}
H= -p_\Omega^2+p_\phi^2+2ce^{6\om+\phi}+\lambda^2e^{-2\phi}= 0,\label{pop}
\end{equation}
which corresponds to a flat universe with dilaton field $\phi$ and
non vanishing antisymmetric field $B_{\mu\nu}$ coming from the
$NS$--$NS$ sector of effective string theory; the existence of this field is reflected only in the positive definite constant $\lambda^2$ (see Refs. \cite{ga00,gave}). Because we are interested in problems of the formalism only, we shall consider $c$ as an arbitrary constant. In  the
case $c<0$ the potential can vanish, so that the system  does not admit an
intrinsic time.  In the
low energy string theory one can consider the limit $\phi\to
-\infty$, then the $e^\phi\equiv V(\phi)$ factor in the first term
of the potential verifies $V(\phi)=V'(\phi)\ll 1$, and we can
replace $ce^\phi$ by the constant $\overline c$ fulfilling
$|\overline c|\ll |c|$:
\begin{equation}
H= -p_\Omega^2+p_\phi^2+2\overline c e^{6\om}+\lambda^2e^{-2\phi}= 0.\label{popy}
\end{equation}
Following Ref. \cite{mos98}, the two-component formulation could now be followed starting from this constraint, after identifying the logarithm of the scale factor, $\Omega$, as time parameter. We shall see that, as it was the case for the closed FRW model in that work,  in the case $\overline c<0$  this choice leads to imaginary eigenvalues. 
For the constraint (\ref{popy}), the application of the formalism of Ref. \cite{mos98} gives
\be
\hat D=-\frac{\partial^2}{\partial\phi^2}+2\overline c e^{6\om}+\lambda^2e^{-2\phi}
\ee
and the corresponding eigenvalue equation
\be
\left\{-\frac{\partial^2}{\partial\phi^2}+2\overline c e^{6\om}+\lambda^2e^{-2\phi}\right\}\Phi_n=E_n^2\Phi_n.
\ee
Then defining
$
{\cal E}_n\equiv E_n^2-2\overline c e^{6\om}\label{im}
$
we obtain
\be
\left\{-\frac{\partial^2}{\partial\phi^2}+\lambda^2e^{-2\phi}\right\}\Phi_n={\cal E}_n\Phi_n.
\ee
We see that, though choosing, as before, ${\cal E}_n\in\mathbb{R}$, in the case $\overline c<0$ the eigenvalues $E_n= \sqrt{{\cal E}_n+2\overline c e^{6\om}}$ corresponding to the operator $\hat H$ (see Eq. (\ref{hal})) become  imaginary for large enough values of the scale factor, as it happens in Ref. \cite{mos98} in the case of the closed ($k=1$) model. Note that this is in correspondence with the fact that the classical quantity from which the operator $\hat D$ comes is $p_\phi^2+2\overline c e^{6\om}+\lambda^2e^{-2\phi}$, which is {\it not} positive definite. But the operator $\hat D$ is defined starting from the previous identification of the time variable.
Thus the existence of imaginary eigenvalues for the operator $\hat H$, in this sense, reflects  that the variable chosen  as time is not globally monotonous:
 Indeed, it is easy to see that $[\Omega,H]$ vanishes for finite values of the canonical variables (the same happens with the dilaton $\phi$): $[\Omega,H]\sim -p_\Omega$, and $p_\Omega$ passes continuously from positive to negative values, so that the scale factor stops growing and begins to decrease. Then there would exist an infinite range of values of ``time'' which are never reached; this range depends on the values of the other variables, i.e. $\phi$ and $p_\phi$, which is in correspondence with the fact that the range of values of $\Omega$ such that the eigenvalues become imaginary also depends on the dilaton.\\

In a description with a globally right notion of time, for $\overline c<0$ the scale factor $\Omega$ cannot be taken as the clock for the system. A right choice becomes apparent by performing the canonical
transformation first introduced for the Taub universe in Ref.
\cite{cafe01} in order to obtain a constraint with only one term
in the potential. This is achieved by introducing the generating
function of the first kind
$
f_1 (\phi,s)= \pm|\lambda|e^{-\phi} \sinh s .$
The new canonical variables are then given by
$
s  =  \pm \mathrm{arcsinh}\left(p_\phi  e^{\phi}/|\lambda|\right),$
 $p_s  = \pm |\lambda|e^{-\phi} \cosh s.$
With this canonical transformation the resulting form for the
Hamiltonian constraint in the limit $V(\phi)=V'(\phi)\ll 1$ is \cite{si03}
\begin{equation}
H=-p_\om^2+p_s^2+2\overline c e^{6\om}=0,\label{joi}
\end{equation}
and we can  apply the two-component formulation starting from this constraint. The central difference with a straightforward application of the formalism to the original constraint is that now the constraint surface splits into two disjoint sheets, identified by the sign of the momentum which does not vanish. The formalism thus includes two disjoint theories for the physical degree of freedom evolving with a globally right time. If $\overline c>0$, we have $p_\Omega\neq 0$ and $\Omega$ is a global time. But in
the case  $\overline c<0$, the momentum  which does not vanish is $p_s$ and
the time is $t=\pm s$. According to the definition of the new
variable $s$, the time $t=\pm s$ is a function of both $p_\phi$ and
$\phi$, then being effectively an {\it extrinsic time}.
The application of the two-component formalism starting from the constraint (\ref{joi}) leads  to different eigenvalue equations depending on the sign of   $\overline c$. In the case $\overline c>0$ (which in fact does not require the canonical transformation), the global time is the scale factor, and we would obtain
\be
\left\{-\frac{\partial^2}{\partial s^2}\right\}\Phi_n=E_n^2\Phi_n,\label{omegat}
\ee
that is, the eigenvalue equation $\hat D\Phi_n={\cal E}_n\Phi_n$ with the time-dependent eigenvalues $E_n=\sqrt{{\cal E}_n+2\overline c e^{6\om}}$, which are kept real as the system evolves.  In the case  $\overline c<0$ requiring an extrinsic time, instead, the eigenvalue equation $\hat D\Phi_n={\cal E}_n\Phi_n$ turns to be 
\be
\left\{-\frac{\partial^2}{\partial\Omega^2}-2\overline c e^{6\om}\right\}\Phi_n=E_n^2\Phi_n,\label{enfi}
\ee
with ${\cal E}_n=E_n^2$, so that now the eigenvalues associated to the operator $\hat H$ are real in a trivial sense, as they do not change when the system evolves. In this case, differing from the situation with $\overline c<0$ before the canonical transformation, the operator $\hat D$ is the quantum version of the classical quantity $p_\Omega^2-2\overline c e^{6\om}$, which for $\overline c<0$ {\it is} positive definite; thus no imaginary eigenvalues would be expected to appear. We emphasize this point: when a right time is identified as a previous step before applying the quantization formalism, the operator $\hat D$ is associated to a classical quantity which cannot be negative nor zero for finite values of the canonical variables.\\

To obtain a better insight it can be helpful a comparison with the usual canonical quantization formalism, that is with the Wheeler--DeWitt or the Schr\"{o}dinger formulations. As early pointed in Ref. \cite{isham}, both formulations are equivalent when the time does not appear in the potential. For the models considered (see (\ref{popy})), the problem of time which we have addressed appears in the case $\overline c <0$, for which, after the canonical transformation introduced above, one obtains a time-independent potential. Thus in this case both the Wheeler--deWitt and the Schr\"{o}dinger formulations are equivalent (in the strong sense that the solutions are exactly the same). Choosing the  Wheeler--deWitt equation we obtain: 
\be
\left[-\frac{\partial^{2}}{\partial\Omega^{2}}+\frac{\partial^{2}}{\partial s^{2}}-2\bar{c}e^{6\Omega}\right]\Psi=0
\ee
In terms of these new variables, the equation is separable. Proposing  the ansatz $\Psi(\Omega,s)=\Theta(\Omega)S(s)$
we  obtain the two ordinary differential equations
\begin{eqnarray}
\frac{d^2\Theta(\Omega)}{d\Omega^{2}}+2\bar{c}e^{6\Omega}\Theta(\Omega)-\kappa\Theta(\Omega) & = & 0\\
\frac{d^{2}S(s)}{ds^{2}}-\kappa S(s) & = & 0
\end{eqnarray}
where $\kappa$ is a separation constant. Unitarity of the  evolution leads  to the choice $\kappa=-E^{2}.$ Then
$
S(s)\sim e^{iEs}
$
and we recover Eq. (\ref{enfi}) for the spatial part of the wave function 
\be
\frac{d^2\Theta(\Omega)}{d\Omega^{2}}+2\bar{c}e^{6\Omega}\Theta(\Omega)+E^{2}\Theta(\Omega)=0.
\ee
Note that the spatial part of the  Wheeler--DeWitt wave function coincides with the function $\Phi$ in this particular case but this is not a general feature. For example, this will not be true for ${\bar c}>0$. In this case  a globally good time is $\Omega$  and the formalism of Ref. \cite{mos98} presents no eigenvalue problems. 
However, the usual canonical quantization procedure  itself is not so easy to interpret. In this case, as a result of the time-dependent potential, the  Wheeler--DeWitt formalism is not equivalent to the Schr\"{o}dinger formalism, thus making not trivial an interpretation in terms of conserved probabilities. In the Schr\"{o}dinger picture the wave function  for the physical degrees of freedom, say $x$, at a time $t$ is given by evolving  the solution  at an initial time $t_0$, which results from the equation  \cite{isham,ss04}: 
 \be
\hat h^2(x,t_0)\Psi_E(x,t_0)=E^2\Psi_E(x,t_0).\label{hath}
\ee 
 The eigenvalue equation $\hat D\Phi_n=E_n^2\Phi_n$ appears similar to this equation; in fact, the operator $\hat h^2$ has the same form of the  operator $\hat D$. But while in the Schr\"{o}dinger case the eigenvalues are constant --the corresponding eigenvalue equation (\ref{hath}) is solved at a fixed time $t_0$-- and the evolution is straightforwardly obtained by acting on $\Psi_E(x,t_0)$ with the evolution operator given by the integral of the time-dependent true Hamiltonian, in the two-component formulation time-dependent eigenvalues appear, and the evolution is not so immediately obtained.\\

We have then focused on a specific problem that arises in the two-component formalism introduced in Ref. \cite{mos98}:  the appearance of imaginary eigenvalues. As pointed in the cited work, this leads to a decomposition of the Hilbert space into three parts: negative norm vectors, positive norm vectors, and null vectors, being the latter associated to imaginary eigenvalues; this decomposition has the peculiarity of being variable, according to the evolution of the corresponding eigenvalues themselves.  
We have shown by means of a specific simple example that this feature can be seen to arise from a choice of time which is not globally right. In other words, we could say that the old but fundamental problem of time in quantum cosmology appears in another disguise. However, we have also showed within an analysis which could, in principle, be extended beyond the particular models considered here, that  a good choice of time  renders the proposed formalism free of the peculiarity of eigenvalues which evolve to become imaginary.

\end{document}